\documentclass[letterpaper]{article} % DO NOT CHANGE THIS
\usepackage{aaai24}  % DO NOT CHANGE THIS
\usepackage{times}  % DO NOT CHANGE THIS
\usepackage{helvet}  % DO NOT CHANGE THIS
\usepackage{courier}  % DO NOT CHANGE THIS
\usepackage[hyphens]{url}  % DO NOT CHANGE THIS
\usepackage{graphicx} % DO NOT CHANGE THIS
\usepackage{natbib}  % DO NOT CHANGE THIS AND DO NOT ADD ANY OPTIONS TO IT
\usepackage{caption} % DO NOT CHANGE THIS AND DO NOT ADD ANY OPTIONS TO IT
\usepackage{algorithm}
\usepackage{algorithmic}
\usepackage{tcolorbox}
\usepackage{booktabs}
\usepackage{lipsum}

\usepackage{xcolor}

\usepackage{newfloat}
\usepackage{listings}
\DeclareCaptionStyle{ruled}{labelfont=normalfont,labelsep=colon,strut=off} % DO NOT CHANGE THIS
\floatstyle{ruled}
\newfloat{listing}{tb}{lst}{}
\floatname{listing}{Listing}
\title{Prompt Refinement or Fine-tuning?\\Best Practices for using LLMs in Computational Social Science Tasks}
\author {
    Anders Giovanni Møller\textsuperscript{\rm 1},
    Luca Maria Aiello\textsuperscript{\rm 1,2},
}
\affiliations {
    \textsuperscript{\rm 1} IT University of Copenhagen, \textsuperscript{\rm 2} Pioneer Center for AI\\
    agmo@itu.dk, 
    luai@itu.dk
}

\begin{document}

\maketitle

\begin{abstract}
Large Language Models are expressive tools that enable complex tasks of text understanding within Computational Social Science. Their versatility, while beneficial, poses a barrier for establishing standardized best practices within the field. To bring clarity on the values of different strategies, we present an overview of the performance of modern LLM-based classification methods on a benchmark of 23 social knowledge tasks. Our results point to three best practices: select models with larger vocabulary and pre-training corpora; avoid simple zero-shot in favor of AI-enhanced prompting; fine-tune on task-specific data, and consider more complex forms instruction-tuning on multiple datasets only when only training data is more abundant.
\end{abstract}

\section{Introduction}

The release of ChatGPT in November 2022 has sparked broad interest for Large Language Models (LLMs) due to their capability to solve complex tasks of text understanding and generation~\cite{bubeck_sparks_2023}. The Computational Social Science (CSS) community has rapidly recognized the potential of LLMs as tools for capturing textual dimensions of semantics and pragmatics -- crucial elements of online discourse that have traditionally been challenging to quantify~\cite{bail2024generative}. %ziems_can_2023

This new opportunity, however, comes with the hurdle of choosing the appropriate use of LLMs in a rapidly-expanding landscape of models and solutions. Prior to the widespread adoption of LLMs, CSS practitioners typically relied on fine-tuning smaller encoder-based models for domain-specific classification tasks~\cite{sun2019fine}. By contrast, LLMs can be used more flexibly, enabling a variety of alternative classification approaches~\cite{chae2023large}. Such versatility, while beneficial, poses a barrier for establishing standardized best practices within the field.

In their most straightforward usage, LLMs can function as zero-shot classifiers, requiring only some target text and a classification prompt~\cite{kojima2022large}. This approach is convenient because it applies the base model without the need of additional training to alter its weights. The prompt can be improved through various strategies, such as manual prompt engineering~\cite{white2023prompt}, automated prompt generation based on the task descriptions~\cite{shin2020autoprompt}, or prompt augmentation with additional task-specific information~\cite{brown2020language} or by integration of external knowledge bases~\cite{li2022survey}.

Alongside approaches that require no training, fine-tuning on domain-specific data may offer better adaptability to specific tasks, albeit at the expense of increased computation~\cite{wei_finetuned_2022}. Instruction-tuning is another flavor of fine-tuning, where the model is conditioned to adhere to explicit instructions and align with human judgments, although crafting high-quality instructions can be both costly and time consuming~\cite{ouyang2022training}. Furthermore, the continuous introduction of new language models raises questions around the effectiveness of different prompting and training techniques across various models.

To bring some clarity on the value of these different practices in the typical workflow of text classification for the Computational Social Sciences, we provide an overview of how current LLM-based methods perform on a variety of CSS text classification tasks. Our goal is to provide practitioners with actionable guidelines on how to prioritize the use of different classification techniques. Specifically, we seek to answer three questions to investigate the effectiveness of the three main families of LLM-based classification:

\vspace{1mm}
\noindent\textbf{RQ1:} What is the value of \emph{prompt-improvement strategies} that add task-relevant knowledge?

\vspace{1mm}
\noindent\textbf{RQ2:} How does \emph{fine-tuning} on static instructions compare with LLM-generated instructions?

\vspace{1mm}
\noindent\textbf{RQ3:} To what extent an increase in the \emph{volume of pre-training data} (e.g., Llama-2 vs. Llama-3) enhances downstream performance?
\vspace{1mm}

We apply 6 state-of-the-art methods on two LLMs and test them against a standard benchmark of $23$ text classification tasks typical of the CSS domain~\cite{choi_llms_2023}. While not fully exhaustive of all possible nuances of classification methods and tasks, our experiments cover the main state-of-the-art classification techniques, with the main goal of providing pragmatic guidelines to practitioners in the field.

\section{Materials and Methods}

We run all our experiments on two open-source models of the Llama series: \texttt{Llama-2-7B-chat} and \texttt{Meta-Llama-3-8B-Instruct}, both released under commercial user license (\url{https://ai.meta.com/llama/license/}). We initialize both models with a temperature value of $0.9$, in line with the setup of previous work. Llama-3 is trained on a corpus of 15T tokens, about seven times larger than Llama-2, and it features a vocabulary size that is four times larger (128K tokens).

\subsubsection{The SOCKET benchmark}

The \textbf{Soc}ial \textbf{K}nowledge \textbf{E}valuation \textbf{T}ests (SOCKET) is a collection of 58 datasets in the domain of social knowledge that can be used to benchmark algorithms for natural language understanding~\cite{choi_llms_2023}. It is the first collective benchmark that has been used to test the capabilities of LLMs in various \emph{social} contexts. The datasets are grouped into five types of task: \emph{humor \& sarcasm}, \emph{offensiveness}, \emph{sentiment \& emotion}, \emph{trustworthiness}, and \emph{social factors}. In addition to the labeled texts, SOCKET provides one prompt for each of the tasks.

In our experiments, we only consider the 44 datasets that refer to classification tasks, saving regression, pair-wise comparisons, and span identification tasks for future work. For fine-tuning, we use the data corresponding to the 44 classification task. For evaluation, we use a representative subset of 23 datasets. We use the same train-test split as defined in~\citet{choi_llms_2023}. To manage computational resources effectively, we constrained our test sample size to up to $2,000$ random samples per task. 

\subsubsection{Zero-shot prompts}

We evaluate the performance of the models using the zero-shot prompts provided in SOCKET (cf.~Prompt 1 in Appendix). The prompts are manually designed and do not include any examples, directing the model to solve tasks without any specific guidance. In this setting, we rely entirely on the LLM's internal representation and understanding of the individual tasks.

\subsubsection{AI-knowledge prompts}

We produce AI-based enhancement to the zero-shot prompts using \emph{generated knowledge prompting}, a technique that relies on a language model to generate task-specific knowledge that can then be used as additional information to be included into the prompt~\cite{liu_generated_2022}. We use GPT-4 to generate task-specific label descriptions based on the zero-shot prompts and the available label options (cf. Prompt~2). This process adds task-aware elements to the prompts, providing descriptions for each individual label-option (cf. Prompt~3). 
%\cite{openai_gpt-4_2023}

\subsubsection{Retrieval-Augmented Generation}

Retrieval-Augmented Generation (RAG) integrates an information retrieval module within the generative framework of a Large Language Model~\cite{lewis_retrieval-augmented_2021}. The RAG system uses the prompt as a query to search a domain-specific knowledge base, retrieving information that is relevant to both the prompt and the domain. The retrieved data is combined with the initial prompt and submitted to the LLM for generation. This methodology is designed to adapt the LLM's output to the target domain without the need of additional training on specialized data~\cite{hu_rag_2024}. Recent empirical studies indicate that RAG presents a competitive alternative to traditional fine-tuning, particularly due to the minimal computational resources required for generating and querying the search index~\cite{balaguer_rag_2024}.

For each task, we apply the \texttt{all-MiniLM-l6-v2} model to create dense embeddings of all the training instances. These vector representations are constructed using text segments of 1000 characters, with a 150-character overlap. We efficiently index all the embeddings with the FAISS library~\cite{douze2024faiss}. During the evaluation phase on the test dataset, we calculate the embedding of the input text, and use it to query the index and retrieve the top five most similar texts, based on cosine similarity, along with their corresponding labels. We then formulate a final prompt for classification, integrating the test sample and the retrieved documents (cf. Prompt~4). In the system prompt, in addition to the specifics of our RAG configuration, we also include the AI-generated descriptions of the labels.

\subsubsection{Fine-tuning}

When Supervised Fine Tuning (SFT) an LLM for a specific classification task, the model is provided with a series of prompts containing: \emph{i)} fixed classification instructions specific to the task, including all the classification labels allowed, and \emph{ii)} a set of labeled texts (cf. Prompt~5). The loss calculated between the generated output and the examples' true labels is used to update the model's weights.

We adopt a two-phase fine-tuning approach that aligns  with the current best practices. During the first phase, we use Quantized Low-Rank Adaptation (QLoRA), an efficient fine-tuning technique~\cite{dettmers_qlora_2023}. In QLoRA, the main model is frozen and quantized to a 4-bit representation. The fine-tuning process is used to learn separate low-rank matrices of gradients, which are then combined with the frozen model during inference, weighted by a factor $\alpha$.

In the second phase, we perform Direct Preference Optimization (DPO), a technique that updates the model's weights based on the explicit user preference for one training example over another~\cite{rafailov_direct_2023}. During DPO, the model receives a prompt and pairs of responses ranked by preference. Based on cross-entropy loss, the model updates its weights to maximize the probability of generating the preferred example.

We trained both phases for one epoch, and we set $\alpha$ to 16, the dropout rate to 0.05, and the matrix rank to 8.

\subsubsection{Instruction tuning}

Instruction tuning is a special type of fine-tuning that, instead of fine-tuning on labeled text examples from a single task, provides the model with a set of instructions and desired corresponding outputs from multiple tasks~\cite{wei_finetuned_2022}. Unlike traditional fine tuning, instruction tuning improves the model's ability to follow classification instructions correctly, and it produces a final model that can be flexibly employed to solve a series of classification tasks within the same domain. To implement instruction tuning, we use the same SFT+DPO pipeline that we employed for fine-tuning, but using instructions and examples from all the tasks during training. This approach is similar to that used by \emph{SocialiteLlama}, the first example of instruction-tuned model on all the tasks from the SOCKET benchmark~\cite{dey_socialite-llama_2024}.

\begin{table*}[t!]
\resizebox{\textwidth}{!}{%
\begin{tabular}{@{}l|cccccc|cccccc@{}}
\toprule
\multicolumn{1}{c|}{\textbf{Tasks}} &
  \multicolumn{6}{c|}{\textbf{Llama-2 7B chat}} &
  \multicolumn{6}{c}{\textbf{Llama-3 8B Instruct}} \\ \cmidrule(l){2-13} 
 &
  Zero-shot &
  AI Knowledge &
  RAG &
  Fine-tuning &
  \begin{tabular}[c]{@{}c@{}}Instruction\\ tuning\end{tabular} &
  \begin{tabular}[c]{@{}c@{}}Reverse\\ Instructions\end{tabular} &
  Zero-shot &
  AI Knowledge &
  RAG &
  Fine-tuning &
  \begin{tabular}[c]{@{}c@{}}Instruction\\ tuning\end{tabular} &
  \begin{tabular}[c]{@{}c@{}}Reverse\\ Instructions\end{tabular} \\ \midrule
\multicolumn{1}{c|}{\textbf{Humor \& Sarcasm}}     &        &       &        &       &        &       &        &        &        &        &        \\
hahackathon\#is\_humor                    & 0.459 & 0.56   & 0.462 & \textbf{0.834} & 0.564  & 0.548 & 0.765  & 0.864  & 0.636 & 0.442 & 0.904  & \textbf{0.933}  \\
sarc                                      & 0.400 & \textbf{0.492} & 0.451 & 0.303 & 0.475 & 0.216 & 0.511 & 0.591 & 0.534 & \textbf{0.689} & 0.499 & 0.628 \\
tweet\_irony                              & 0.313 & \textbf{0.497}  & 0.366 & 0.458 & 0.464  & 0.638 & 0.540  & 0.663  & 0.551 & 0.510 & \textbf{0.889}  & 0.788  \\ \midrule
\multicolumn{1}{c|}{\textbf{Offensiveness}}        &        &       &        &       &        &       &        &        &        &        &        \\
contextual-abuse\#PersonDirectedAbuse     & 0.103 & 0.480  & 0.182 & \textbf{0.990} & 0.105  & 0.052 & 0.671  & 0.655  & 0.460 & 0.975 & \textbf{0.992}  & 0.978  \\
implicit-hate\#explicit\_hate             & 0.090 & 0.142  & 0.123 & 0.788 & 0.139  & \textbf{0.799} & 0.665  & 0.517  & 0.447 & 0.950 & \textbf{0.951}  & 0.947  \\
contextual-abuse\#IdentityDirectedAbuse   & 0.076 & 0.515  & 0.255 & \textbf{0.883} & 0.102  & 0.001 & 0.708  & 0.758  & 0.516 & 0.893 & \textbf{0.984}  & 0.973  \\
hasbiasedimplication                      & 0.245 & 0.426  & 0.574 & 0.530 & 0.390  & \textbf{0.767} & 0.463  & 0.499 & 0.432 & 0.487 & 0.577  & \textbf{0.833}  \\
hateoffensive                             & 0.503 & 0.326  & 0.625 & 0.765 & 0.548  & \textbf{0.776} & 0.488  & 0.424  & 0.440 & \textbf{0.870} & 0.838  & 0.841  \\
intentyn                                  & 0.090 & 0.157  & 0.463 & 0.158 & 0.251  & \textbf{0.595} & 0.566  & 0.289  & 0.261 & 0.413 & 0.719  & \textbf{0.741}  \\
tweet\_offensive                          & 0.412 & 0.577  & 0.723 & \textbf{0.762} & 0.533  & 0.506 & 0.693  & 0.702  & 0.698 & \textbf{0.837} & 0.822  & 0.688  \\
implicit-hate\#implicit\_hate             & 0.085 & 0.202  & 0.108 & 0.449 & 0.268  & \textbf{0.466} & 0.589  & 0.494  & 0.45 & \textbf{0.783} & 0.762  & 0.737  \\
implicit-hate\#stereotypical\_hate        & 0.047 & 0.164  & 0.725 & \textbf{0.892} & 0.150   & 0.769 & 0.329  & 0.499  & 0.378 & 0.887 & \textbf{0.953}  & 0.929  \\ \midrule
\multicolumn{1}{c|}{\textbf{Sentiment \& Emotion}} &        &       &        &       &        &       &        &        &        &        &        \\
empathy\#distress\_bin                    & 0.048 & \textbf{0.565}  & 0.554 & 0.349 & 0.172  & 0.494 & 0.285  & 0.597  & \textbf{0.667} & 0.382 & 0.602  & 0.500    \\
dailydialog                               & 0.167 & 0.561  & 0.107 & 0.253 & 0.154  & \textbf{0.782} & 0.382  & 0.336  & 0.109 & \textbf{0.839} & 0.837  & 0.655  \\
tweet\_emotion                            & 0.450 & 0.623  & \textbf{0.680} & 0.650 & 0.498  & 0.319 & 0.725  & 0.776  & 0.771 & \textbf{0.802} & 0.721  & 0.750  \\
crowdflower                               & 0.215 & 0.288  & 0.224 & \textbf{0.303} & 0.235  & 0.154 & 0.179  & 0.243  & 0.282 & 0.342 & 0.286  & \textbf{0.353}  \\ \midrule
\multicolumn{1}{c|}{\textbf{Social Factors}}       &        &       &        &       &        &       &        &        &        &        &        \\
hayati\_politeness                        & 0.281 & 0.438  & \textbf{0.688} & 0.500 & 0.375  & 0.25  & 0.844  & 0.656  & 0.656 & 0.719 & \textbf{0.844}  & 0.688  \\
complaints                                & 0.438 & 0.649  & 0.780 & \textbf{0.901} & 0.562  & 0.559 & 0.806  & 0.878  & 0.809 & \textbf{0.916} & 0.872  & 0.817  \\
stanfordpoliteness                        & 0.550 & 0.621  & \textbf{0.665} & 0.522 & 0.582  & 0.439 & 0.640  & 0.644  & 0.621 & \textbf{0.678} & 0.549  & 0.550  \\
questionintimacy                          & 0.155 & \textbf{0.222}  & 0.204 & 0.209 & 0.227  & 0.182 & 0.2    & 0.204  & 0.2    & 0.320 & \textbf{0.351}  & 0.347  \\ \midrule
\multicolumn{1}{c|}{\textbf{Trustworthyness}}      &        &       &        &       &        &       &        &        &        &        &        \\
hypo-l                                    & 0.269 & 0.402  & 0.557 & 0.437 & 0.349  & 0.672 & 0.665  & 0.693  & 0.536 & \textbf{0.724} & 0.712  & 0.721  \\
rumor\#rumor\_bool                        & 0.282 & 0.606  & \textbf{0.887} & 0.444 & 0.458  & 0.592 & 0.514  & 0.542  & 0.549 & 0.620 & 0.647  & \textbf{0.669}  \\
two-to-lie\#receiver\_truth               & 0.490 & 0.430  & 0.899 & \textbf{0.945} & 0.549  & 0.449 & 0.366  & 0.613  & 0.682 & \textbf{0.945} & 0.943  & 0.933  \\ \midrule
\multicolumn{1}{c|}{\textbf{Cross-task average}} & 0.268 & 0.432 & 0.491 & \textbf{0.579} & 0.354 & 0.479 & 0.547 & 0.571 & 0.508 & 0.697 & \textbf{0.750} & 0.739 \\ \bottomrule
\end{tabular}%
}
\caption{Accuracy on SOCKET classification tasks across models. Best results for each model are highlighted in bold.}
\label{tab:full-results}
\end{table*}

\subsubsection{Reverse instruction tuning}

Instruction tuning typically relies on one fixed human-generated instruction for each task. This constrains the ability of LLMs to learn associations between the semantics of instructions and their corresponding responses. Generating synthetic instruction variants with LLMs mitigates this problem without needing extensive human labor~\cite{moller_parrot_2024}. This process is known as \emph{reverse instruction generation}~\cite{koksal_longform_2024}. It involves presenting the LLM with a textual output and prompting it to formulate a plausible instruction that could lead to that output (cf. Prompt~6). We extend this method to create instructions that are specific to classification tasks consisting of a target text, a set of possible labels, and the label for the given text (cf. Prompt~7).

For generating reverse instructions, we randomly sample up to $4,000$ samples from each task's training set. We use OpenAI's \texttt{gpt-3.5-turbo-0125} as LLM, setting its temperature to 1, to ensure the generation of diverse instructions. For each task, we try generate up to $4,000$ new instructions for training, and 400 for each validation and test. In total, we generate $179,510$ samples. We then clean the output using simple heuristics designed to remove noisy generations, filtering out instructions that repeat the input text, explicitly reveal the label, or are improperly formatted.  We create a new training set for instruction-tuning by simply replicating each training example for all its instruction variants, and then apply the SFT+DPO pipeline. During the evaluation phase, we randomly sample instructions from the training set and integrate them into the prompt template.

\section{Results}

Table~\ref{tab:full-results} presents the classification accuracy across methods and tasks. A critical factor impacting performance is the selection of the pre-trained model. On average, across tasks, there is an accuracy improvement ranging from $0.02$ to $0.4$ when employing Llama-3 over Llama-2 \textbf{(RQ3)}. This result indicates that there is still room for improving the the language models' understanding during pre-training, and suggests that switching to recent models is worth prioritizing. 

When comparing the performance of prompt enhancement methods, two main findings emerge. First, zero-shot yields relatively high accuracy, yet it is consistently outperformed by AI-generated knowledge prompting \textbf{(RQ1)}. This trend is not as pronounced in the \emph{offensiveness} category, where some tasks exhibit a notable decrease in accuracy with AI-enhanced prompts. This could be attributed to the safeguards built into the LLMs when addressing sensitive content, potentially restricting their ability to generate high-quality prompts. Second, the performance of Retrieval-Augmented Generation (RAG) for prompt enhancement is inconsistent. Its relative performance to zero-shot is generally better with Llama-2, albeit with considerable variability across tasks, and tends to be less effective with Llama-3 \textbf{(RQ1)}. This suggests that models with less extensive pre-training may benefit from external knowledge integration, but this advantage diminishes with models that have a more robust pre-training foundation.

Fine-tuning markedly improves the accuracy of AI-knowledge prompting by an average of $0.15$ with Llama-2 and $0.13$ with Llama-3. In contrast to traditional fine-tuning, which directly modifies model weights, parameter-efficient fine-tuning using QLoRA is less resource-demanding and achieves good results with relatively small training sets, making it a practical alternative in many scenarios. The two forms of instruction tuning, however, yield divergent outcomes depending on the model. Llama-2's performance declines by an average of $0.22$ with instruction tuning and by $0.1$ with reverse instruction tuning, with many tasks experiencing accuracy drops even greater than $0.3$. Conversely, Llama-3 shows a modest increase in accuracy of approximately $0.05$ on average. This disparity may be due to Llama-3's superior capability to process complex and semantically varied input data, thanks to its expanded vocabulary and training corpus. The added complexity, however, introduces noise into Llama-2's classification process, suggesting a need for more fine-tuning data to bridge the performance gap with Llama-3. In summary, the results indicate that advanced fine-tuning methods involving small sets of instructions and data from multiple tasks hold some promise but also risk performance decline if the foundational model lacks the necessary expressiveness \textbf{(RQ2)}. Moreover, while reverse instructions enhance training diversity, they can also lead to hallucinations and information leaks that require manual intervention, thus limiting their practicality.

The robustness of our findings is supported by the limited performance variation across task categories, which can be largely attributed to the difficulty of individual tasks. For instance, the \emph{crowdflower} task exhibits the lowest performance due to its 13 possible classes that represent concepts challenging to discern from textual information.

\section{Conclusion}

Our findings highlight three good practices that practitioners can adopt when using LLMs for classification tasks within the field of Computational Social Science. First, the selection of the model is a crucial decision that significantly impacts performance. Choosing models that have undergone extensive pre-training is recommended. Second, basic zero-shot methods should be avoided in favor of enhanced zero-shot techniques that incorporate LLM-generated descriptions of the task and labels  into the prompt. This straightforward method offers substantial benefits relative to its minimal cost, unlike more complex retrieval-based methods for prompt augmentation, which do not appear as effective for classification purposes. Last, fine-tuning should be pursued whenever adequate computational resources are accessible, as it consistently yields positive results and can be executed cost-effectively using contemporary methods like QLoRa. Nevertheless, in scenarios where fine-tuning data is scarce, advanced instruction tuning that integrates instructions and datasets from diverse tasks should be approached with caution, as it may not generalize well and could potentially degrade performance.

\clearpage

\section*{Appendix}

%%%% PROMPT 1
\begin{tcolorbox}[float, floatplacement=h!,colback=green!5,colframe=green!40!black,coltitle=white,title=\textbf{Prompt 1}: Zero-shot prompt]
{\scriptsize
\texttt{\textit{\# System prompt}}

\texttt{You are a helpful, respectful and honest assistant.}\\

\texttt{\textit{\# Task prompt (example for the} sarc \textit{task)}}

\texttt{For the sentence: \{text\}, is it sarcastic?}\\

\texttt{You can choose from the following labels: \{labels\}.}\\

\texttt{Answer:}
}
\end{tcolorbox}

%%%% PROMPT 2
\begin{tcolorbox}[float, floatplacement=h!,colback=green!5,colframe=green!40!black,title=\textbf{Prompt 2}: AI-Knowledge Generation]
{\scriptsize
\texttt{\textit{\# Task prompt}}

\texttt{For the task: \{task\_description\}. Explain briefly the labels: \{labels\_list\}}\\
}
\end{tcolorbox}

%%%% PROMPT 3
\begin{tcolorbox}[float, floatplacement=h!,colback=green!5,colframe=green!40!black,title=\textbf{Prompt 3}: Knowledge-improved Zero-shot]
{\scriptsize
\texttt{\textit{\# System prompt}}

\texttt{You are a helpful, respectful and honest assistant. You have the following knowledge about task-specific labels: 'sarcastic': This label indicates the sentence is sarcastic, meaning it conveys irony or mocks with a tone of detachment or insincerity. 'literal': This label is used if the sentence is not sarcastic, implying a straightforward or sincere expression without irony.}\\

\texttt{\textit{\# ... Remainder of prompt as in Prompt 1 ...}}
}
\end{tcolorbox}

%%%% PROMPT 5
\begin{tcolorbox}[float, floatplacement=h!,colback=green!5,colframe=green!40!black,title=\textbf{Prompt 5:} Fine-tuning Prompt]
{\scriptsize
\texttt{\textit{\# System prompt}}

\texttt{You are a helpful, respectful and honest assistant.}\\

\texttt{\textit{\# Task prompt}}

\texttt{For the sentence: \{task\_description\_with\_text\}
You can choose from the following labels: \{label\_list\}.\ Answer: \{label\}
}
}
\end{tcolorbox}

%%%% PROMPT 4
\begin{tcolorbox}[float, floatplacement=h!,colback=green!5,colframe=green!40!black,title=\textbf{Prompt 4:} RAG Prompt]
{\scriptsize
\texttt{\textit{\# System prompt}}

\texttt{You are part of a RAG classification system designed to categorize texts. \textit{Continued specification of the RAG...}}\\

\texttt{\textit{\# Task prompt}}

\texttt{Consider the relevance and content of each document in relation to the input text and the descriptions of the labels. If a retrieved document is highly relevant to the input text and aligns closely with the description of a label, that label might be the correct classification.}\\
    
\texttt{Retrieved Documents:}

\texttt{Document $i$: \{doc $i$\}}\\

\texttt{Input Text: \{text\}}\\
\texttt{Answer: [/INST]}
}
\end{tcolorbox}

%%%% PROMPT 6
\begin{tcolorbox}[float, floatplacement=h!,colback=green!5,colframe=green!40!black,title=\textbf{Prompt 6:} Reverse Instructions Generation Prompt]
{\scriptsize

\texttt{Instruction: X}

\texttt{Output: \{doc\}}

\texttt{What kind of instruction could this be the answer to?}

\texttt{X:}
}
\end{tcolorbox}

%%%% PROMPT 7
\begin{tcolorbox}[float, floatplacement=h!,colback=green!5,colframe=green!40!black,title=\textbf{Prompt 7:} Reverse Instructions Generation for Classification]
{\scriptsize
\texttt{\textit{\# System prompt}}

\texttt{You are a helpful assistant helping in creating instructions for a text classification task.}\\

\texttt{\textit{\# Task prompt}}

\texttt{Instruction: X}\\

\texttt{Input: \{text\}}\\

\texttt{Labels: \{label\_list\}}\\

\texttt{Output: \{label\}}\\

\texttt{What kind of instruction could ``Output'' be the answer to given ``Input'' and ``Labels''? Please make only an instruction for the task and include brief descriptions of the labels.}\\

\texttt{Begin your answer with 'X: '}
}
\end{tcolorbox}

\end{document}